\documentclass[preprint,showpacs,preprintnumbers,nofootinbib]{revtex4}
\usepackage[dvips]{color}
\usepackage{graphicx}

\newcommand{\diag}{{\rm diag}}

\newcommand{\gev}{{\rm GeV}}

\begin{document}

\title{Quark Masses and Mixings with Hierarchical Friedberg-Lee Symmetry}
\author{Takeshi Araki\footnote{araki@phys.nthu.edu.tw}  and 
C.~Q.~Geng\footnote{geng@phys.nthu.edu.tw} }
\affiliation{Department of Physics, National Tsing Hua University,
Hsinchu, Taiwan 300}

\begin{abstract}
We consider the Friedberg-Lee  symmetry for the quark sector and show that 
the symmetry closely relates to both quark masses and mixing angles.
We also extend our scheme to the fourth generation quark model and 
find the relation 
$|V_{tb^{'}}| \simeq |V_{t^{'}b}| \simeq m_b/m_{b^{'}}< \lambda^2$ 
with $\lambda\simeq0.22$ for $m_b=4.2\ \gev$ and $m_{b^{'}}>199\ \gev$.
\end{abstract}


\maketitle

\section{Introduction}
Although the standard model (SM) is a very successful
theory, there are still some mysteries and problems.
One of them is the flavor structure of fermions.
We currently know that quarks mix with each other through the 
Cabibbo-Kobayashi-Maskawa (CKM) matrix \cite{ckm} and that there is 
a hierarchy among the three mixing angles of the CKM matrix: 
$\theta_{12}\simeq \lambda \simeq 0.22$, \ $\theta_{23} \simeq \lambda^2$
and $\theta_{13} \simeq \lambda^4$ \cite{wolf}.
Yet another hierarchy exists among quark masses and it can be expressed
in terms of $\lambda$ as $m_u/m_c \simeq m_c/m_t \simeq \lambda^4$ and
$m_d/m_s \simeq m_s/m_b \simeq \lambda^2$ \cite{mass-mix}.
In particular, $\theta_{12}$ and $\sqrt{m_d/m_s}$ surprisingly coincide 
with each other. 
However, the origin of the hierarchies is still unclear.
This is because the Yukawa sector in the SM contains a huge number of 
unknown parameters.
Consequently, fermion masses and mixing angles remain free parameters in the SM.
Thus, it is interesting to extend the SM with a family symmetry.
In Ref. \cite{FN}, authors introduce a family $U(1)$ symmetry and try 
to explain the quark mass hierarchies and small mixing angles
via hierarchically suppressed nonrenormalizable Yukawa interactions.
A finite group could also be the prime candidate of a family symmetry, 
which can reduce the number of parameters in the Yukawa sector and make the theory
predictive \cite{LM,CM,BK,MST,AK}.

Recently, Friedberg and Lee proposed a (hidden) translational family 
symmetry called Friedberg-Lee (FL) symmetry \cite{FL} 
and tried to relate quark mixing angles with the symmetry.
A more detailed analysis is given in Ref. \cite{ren} and an application to 
the fourth generation quark model is discussed in Ref. \cite{soni}.
Furthermore, many attempts for the lepton sector are performed  
in Refs. \cite{FLlep,LH}, 
some possible origins of the symmetry have been discussed in Ref. \cite{orgn-FL}, 
and a new spontaneous CP violation mechanism with 
the symmetry is considered in Ref. \cite{time}.
Moreover, one of the motivations to consider the FL symmetry is that 
the smallness of up- and down-quark masses as well as the electron mass 
can be naturally understood as it always makes one of three 
generation fermions massless \cite{FL,LH}.
However, the original approach \cite{FL} 
with the FL symmetry cannot explicitly 
reveal the hierarchies of quark masses and mixing angles, which are treated 
as input parameters.
In this paper, we extend the discussion to connect both quark masses and 
mixing angles with the FL symmetry.
In particular, as will be explained in the next section, 
we propose the FL symmetry which translates the three generation 
quarks hierarchically and also show that the hierarchical patterns can be 
responsible for the quark flavor structures mentioned above.

This paper is organized as follows.
In Sec. II, we present the hierarchical FL symmetry.
In Sec. III, we break both FL and CP symmetries in order to generate 
light quark masses and the CP violating Dirac phase in the CKM matrix.
In Sec. IV, we extend our model to the fourth generation case.
We summarize our results in Sec. V.

\section{Hierarchical Friedberg-Lee symmetry}
We start our discussion with the Lagrangian of the quark mass terms 
\begin{eqnarray}
-{\cal L} = M^d_{ij} \overline{d_{i}}d_{j}
 +M^u_{ij} \overline{u_{i}}u_{j}\ ,
\end{eqnarray}
where $M^{u,d}$ are up- and down-type quark mass matrices, which are 
assumed to be symmetric.
The subscripts $i,j=1,...,3$ stand for the family indices.
For the Lagrangian, we impose the FL symmetry \cite{FL}
\begin{eqnarray}
 q_i \rightarrow q_i +(1,\eta_q,\eta_q\xi_q)^T z_q ,\label{eq:FL}
\end{eqnarray}
where $\eta_q$ and $\xi_q$ are c-numbers, $z_q$ is a global Grassmann parameter
and $q=d,u$.
Because of the symmetry, the quark mass matrices take the form
\begin{eqnarray}
M^q=
 \left(\begin{array}{ccc}
 B_q \eta_q^2 + C_q & -B_q \eta_q & -C_q / (\eta_q \xi_q)\\
 -B_q \eta_q & A_q \xi_q^2 + B_q & -A_q \xi_q \\
 -C_q / (\eta_q \xi_q) & -A_q \xi_q & 
 A_q + C_q / (\eta_q \xi_q)^2
 \end{array}\right),
\end{eqnarray}
where we have assumed that $A_q,\ B_q,\ C_q,\ \eta_q,$ and $\xi_q$ are real.
Hence, the theory is CP conserving.
Also, the up and down quarks are massless because of the FL symmetry.
In the next section, we will insert phase factors 
into the mass matrices to generate the light quark masses and CP violation.

In order to make our point more clear, 
here we redefine the parameters as follows:
\begin{eqnarray}
&&C_u \rightarrow C_u \eta_u^2 \xi_u , \\
&&C_d \rightarrow C_d \eta_d^2 \xi_d,\ 
A_d \rightarrow A_d / \xi_d .
\end{eqnarray}
Consequently, the up- and down-type quark mass matrices become
\begin{eqnarray}
M^u=
 \left(\begin{array}{ccc}
 (B_u + C_u \xi_u)\eta_u^2 & -B_u \eta_u & -C_u \eta_u \\
 -B_u \eta_u & A_u \xi_u^2 + B_u & -A_u \xi_u \\
 -C_u \eta_u & -A_u \xi_u & A_u + C_u / \xi_u
 \end{array}\right)
\label{eq:Mu}
\end{eqnarray}
and
\begin{eqnarray}
M^d=
 \left(\begin{array}{ccc}
 (B_d + C_d \xi_d)\eta_d^2 & -B_d \eta_d & -C_d \eta_d \\
 -B_d \eta_d & A_d \xi_d + B_d & -A_d\\
 -C_d \eta_d & -A_d & (A_d + C_d) / \xi_d
 \end{array}\right)
\label{eq:Md} ,
\end{eqnarray}
respectively.
In this basis, we impose
\begin{eqnarray}
A_q \simeq B_q \simeq C_q\ \ {\rm and}\ \ \eta_q, \xi_q \ll 1\ .
\label{eq:para}
\end{eqnarray}
We note that with Eq. (\ref{eq:para}), the down-type quark mass matrix 
becomes similar to the hybrid texture discussed in Ref. \cite{Ryo}.
On the other hand, the up-type one is almost diagonal.
Since $M^{u,d}$ are real and symmetric matrices, they can 
be diagonalized by real orthogonal matrices:
$V_q^T M^q V_q = \diag(m_{q1},m_{q2},m_{q3})$.
With Eq. (\ref{eq:para}), $V_q$ can approximately be written as
\begin{eqnarray}
V_u\simeq
 \left(\begin{array}{ccc}
 1 & -\eta_u & 0 \\
 \eta_u & 1 & 0\\
 0 & 0 & 1
 \end{array}\right) \label{eq:Vu}
\end{eqnarray}
with $m_u/m_c \simeq \eta_u^2$ and $m_c/m_t \simeq |\xi_u|$
for the up-quark sector, and
\begin{eqnarray}
V_d\simeq
 \left(\begin{array}{ccc}
 1 & -\eta_d & -\frac{1}{2}\eta_d\xi_d \\
 \eta_d & 1 & -\frac{1}{2}\xi_d \\
 \eta_d\xi_d & \frac{1}{2}\xi_d & 1
 \end{array}\right)\label{eq:Vd}
\end{eqnarray}
with $m_d/m_s \simeq \eta_d^2$ and $m_s/m_b \simeq 1/2|\xi_d|$
for the down-quark sector, respectively.
Note that the factor $1/2$ in Eq. (\ref{eq:Vd}) arises from 
the $\{33\}$ element of Eq. (\ref{eq:Md}).
From the mass ratios, we can deduce that
\begin{eqnarray}
&&|\eta_u| \simeq \lambda^2 ,\ \ |\xi_u| \simeq \lambda^4 ,\\
&&|\eta_d| \simeq \lambda ,\ \ |\xi_d| \simeq \lambda^2 ,
\end{eqnarray}
with $\lambda\simeq 0.22$.
Therefore, in what follows, we consider the following hierarchical FL translation:
\begin{eqnarray}
&&u_i \rightarrow u_i +(1,-\lambda^2,-\lambda^6)^T z_u\ ,\label{eq:hFLu}\\
&&d_i \rightarrow d_i +(1,\lambda,\lambda^3)^T z_d\ .\label{eq:hFLd}
\end{eqnarray}
The minus signs in the up sector come from $\eta_u$ which is 
negative to reproduce realistic CKM elements.
We note that although the up and down quarks are massless at this stage, 
in the above approximation, we remain $m_u$ and $m_d$ to be nonzero
in order to determine the order of $\eta_q$.
The parameters $\eta_u =-\lambda^2$ and $\eta_d =\lambda$ would be the origin 
of $m_u/m_c \sim \lambda^4$ and $m_d/m_s \sim \lambda^2$, respectively, when
we introduce the symmetry breaking terms.

On the other hand, the parameters $\eta_q$ and $\xi_q$ can also be the 
origin of the tiny mixing angles of the CKM matrix.
For instance, three elements of the CKM matrix can be estimated as
\begin{eqnarray}
&&|V_{us}| \simeq |-\eta_d + \eta_u| \simeq 1.25\lambda ,\nonumber\\
&&|V_{ub}| \simeq 0.5|-\eta_d\xi_d - \eta_u\xi_d |
  \simeq 1.5 \lambda^4 ,\label{eq:estim}\\
&&|V_{cb}| \simeq 0.5|\xi_d | \simeq 0.5\lambda^2 ,\nonumber
\end{eqnarray}
where we have used $\lambda\simeq 4\lambda^2$. 
These results well coincide with the experimental values 
$\theta_{12}\simeq\lambda ,\ \theta_{23}\simeq\lambda^2 ,$ and
$\theta_{13}\simeq\lambda^4$ mentioned in the Introduction. 
Namely, in our model, the flavor structures in the quark sector are explained 
by the hierarchical patterns of the symmetry. 
Hence, it is also easy to establish the relations between the masses and 
mixing angles, such as
\begin{eqnarray}
\sqrt{m_d/m_s} \sim |V_{us}| \label{eq:masmix1}
\end{eqnarray}
via $\eta_d$, and
\begin{eqnarray}
m_s/m_b \sim |V_{cb}| \label{eq:masmix2}
\end{eqnarray}
via $\xi_d$, respectively.

\section{FL symmetry breaking and parameter fitting}
In order to generate the light quark masses, 
the FL symmetry must be broken.
Since we do not know the origin of the breaking, 
there may exist many possible ways to break the symmetry.
Here, we aim at a minimal scheme and 
put phase factors into the mass matrices to break the FL and CP
symmetries simultaneously as discussed in Ref. \cite{FL}.
That is, we replace Eqs. (\ref{eq:Mu}) and (\ref{eq:Md}) as
\begin{eqnarray}
M^u=
 \left(\begin{array}{ccc}
 (B_u + C_u \xi_u)\eta_u^2 & -B_u \eta_u\ e^{i\theta_1} & -C_u \eta_u \\
 -B_u \eta_u\ e^{i\theta_1} & A_u \xi_u^2 + B_u & -A_u \xi_u\ e^{i\theta_2} \\
 -C_u \eta_u & -A_u \xi_u\ e^{i\theta_2} & A_u + C_u / \xi_u
 \end{array}\right),
\label{eq:Mu2}
\end{eqnarray}
\begin{eqnarray}
M^d=
 \left(\begin{array}{ccc}
 (B_d + C_d \xi_d)\eta_d^2 & -B_d \eta_d\ e^{i\phi_1} & -C_d \eta_d \\
 -B_d \eta_d\ e^{i\phi_1} & A_d \xi_d + B_d & -A_d\ e^{i\phi_2} \\
 -C_d \eta_d & -A_d\ e^{i\phi_2} & (A_d + C_d) / \xi_d
 \end{array}\right),
\label{eq:Md2}
\end{eqnarray}
respectively.
Note that phase factors are added into $\{12\},\{21\},\{23\}$ and $\{32\}$ 
elements for each matrix.
The phases $\theta_1$ and $\phi_1$ are responsible for the up- and 
down-quark masses.
If we regard $(e^{i\theta_1}-1)$ and $(e^{i\phi_1}-1)$ as perturbations, 
we obtain
\begin{eqnarray}
 \frac{m_u}{m_c}\simeq \left|2\eta_u^2 (e^{i\theta_1}-1)\right|,\ \ 
 \frac{m_d}{m_s}\simeq \left|2\eta_d^2 (e^{i\phi_1}-1)\right|\ ,
\end{eqnarray}
as we expected.
The phases $\theta_2$ and $\phi_2$ are added to account for the CP violating 
Dirac phase in the CKM matrix.

\begin{figure}[t]
\begin{center}
\includegraphics*[width=0.8\textwidth]{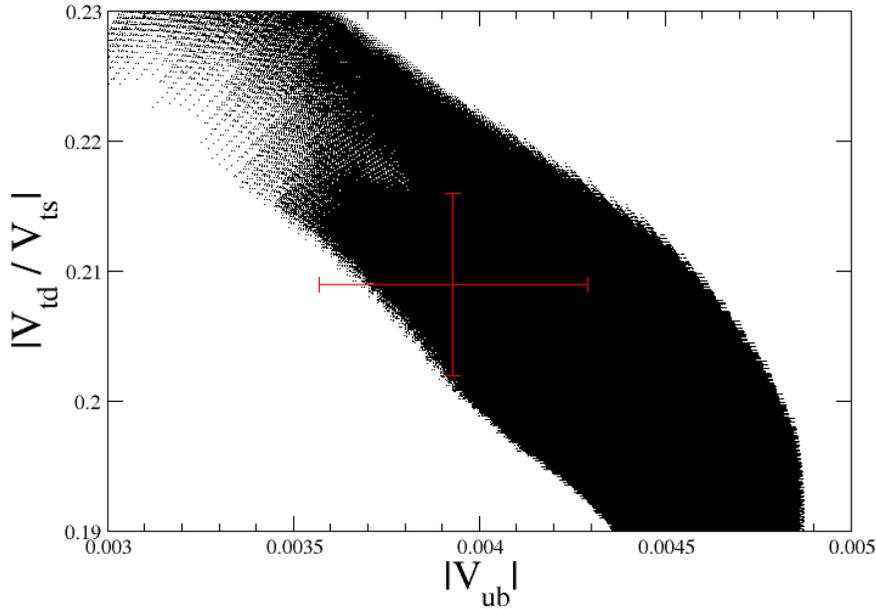}
\caption{\footnotesize
The allowed region in the $|V_{ub}| - |V_{td}/V_{ts}|$ plane, where 
the experimental values from the PDG \cite{PDG} are also plotted 
as the cross lines. 
}\label{fig:Vub-Vtdts}
\end{center}
\end{figure}

In the following, we will present our numerical analysis to illustrate 
our result.
In our calculation, without loss of generality, we will normalize 
Eqs. (\ref{eq:Mu2}) and (\ref{eq:Md2}) with $B_{u,d}=1$.
We will also ignore the terms associated with $A_u$ because they are 
suppressed by $\xi_u = \lambda^4 $.
As a result, the model has three real parameters: 
$A_{d}$ and $C_{u,d}$ , 
and three CP violating phases: $\theta_{1}$ and $\phi_{1,2}$.
For these six parameters, we use the following experimental 
values from the Particle Data Group (PDG) \cite{PDG}:
\begin{eqnarray}
&&m_u / m_c=(0.112 - 0.284)\times 10^{-2} ,\ 
  m_c / m_t=(0.669 - 0.792)\times 10^{-2} ,\\
&&m_s / m_d=17 - 22 ,\ 
  m_s / m_b=(0.160 - 0.315)\times 10^{-1} ,\\
&&|V_{us}|=0.2236 - 0.2274 ,\ 
  |V_{cb}|=0.0401 - 0.0423 ,
\end{eqnarray}
as input parameters.
Figure \ref{fig:Vub-Vtdts} shows the allowed region in the 
$|V_{us}| - |V_{td}/V_{ts}|$ plane with the bounds \cite{PDG}
\begin{eqnarray}
|V_{ub}|=(3.57 - 4.29)\times 10^{-3},\ 
 &&|V_{td}/V_{ts}|=0.202 - 0.216 .
\end{eqnarray}
One can easily see from Fig. \ref{fig:Vub-Vtdts} that our results on 
$|V_{us}|$ and $|V_{td}/V_{ts}|$ are consistent with the experimental data.
Here, we would like to emphasize that the flavor structures of the quark sector 
are mainly originated from the hierarchical FL symmetry 
even though the parameters $A_d$ and $C_{u,d}$ are not completely fixed.
At the best fit point, we have
\begin{eqnarray}
C_u \simeq 0.4,\ A_d\simeq 1.6,\ C_d\simeq 0.34
\end{eqnarray}
with $B_{u,d}=1$, which fill the gap between the exact values and expected ones 
estimated in the previous section.

\section{Implications for fourth generation quark model}
In our scheme, quark masses and mixing angles are closely related with each 
other, such as those in Eqs. (\ref{eq:masmix1}) and (\ref{eq:masmix2}).
So, it should be interesting to extend our scheme to the fourth generation quark 
model and see whether there exist relations between the fourth generation
quark masses and their mixing angles.

We extend the FL symmetry as
\begin{eqnarray}
 q_i \rightarrow q_i +(1,\eta_q,\eta_q\xi_q,\eta_q\xi_q\rho_q)^T z_q\ ,
 \label{eq:FL4}
\end{eqnarray}
and the mass matrices Eqs. (\ref{eq:Mu}) and (\ref{eq:Md}) as
\begin{eqnarray}
M^u&=&
\left(\begin{array}{cccc}
 (B_u+C_u\xi_u+D_u\xi_u\rho_u)\eta_u^2 & -B_u \eta_u & 
   -C_u \eta_u & -D_u\eta_u\\
 -B_u \eta_u & A_u\xi_u^2+B_u+E_u\xi_u^2\rho_u^2 & 
   -A_u \xi_u & -E_u\xi_u\rho_u\\
 -C_u \eta_u & -A_u \xi_u & 
   A_u + \frac{C_u}{\xi_u} +F_u\xi_u^2\rho_u^3 & -F_u\xi_u^2\rho_u^2 \\
 -D_u\eta_u & -E_u\xi_u\rho_u & 
   -F_u\xi_u^2\rho_u^2 & \frac{D_u}{\xi_u\rho_u} + E_u + F_u\xi_u^2\rho_u
\end{array}\right)\nonumber\\  
\label{eq:Mu4}
\end{eqnarray}
and
\begin{eqnarray}
M^d&=&
 \left(\begin{array}{cccc}
 (B_d+C_d\xi_d+D_d\xi_d\rho_d)\eta_d^2 & -B_d \eta_d & 
   -C_d \eta_d & -D_d\eta_d\\
 -B_d \eta_d & A_d\xi_d+B_d+E_d\xi_d\rho_d & -A_d & -E_d\\
 -C_d \eta_d & -A_d & 
   \frac{A_d+C_d}{\xi_d} +F_d\frac{\rho_d}{\xi_d} & -\frac{F_d}{\xi_d} \\
 -D_d\eta_d & -E_d & -\frac{F_d}{\xi_d} & \frac{D_d+E_d+F_d}{\xi_d\rho_d}
\end{array}\right), \nonumber\\  
\label{eq:Md4}
\end{eqnarray}
respectively.
Here, we again assume that
\begin{eqnarray}
A_q \simeq B_q \simeq C_q \simeq D_q \simeq E_q \simeq F_q\ \ {\rm and}
\ \ \eta_q, \xi_q, \rho_q \ll 1\ .\label{eq:para2}
\end{eqnarray}
Note that we extend the model so that the mass matrices keep the features
mentioned just behind Eq. (\ref{eq:para}).
Since a detailed analysis goes beyond the purpose of the paper, 
we would like to roughly study and try to figure out their implications.
Both Eqs. (\ref{eq:Mu4}) and (\ref{eq:Md4}) are diagonalized by 
real orthogonal matrices.
One can easily find that $\rho_q$ determine the mass ratios of the fourth 
and third generation quarks:
\begin{equation}
\frac{m_t}{m_{t^{'}}} \sim \rho_u\ ,\ \ 
\frac{m_b}{m_{b^{'}}} \sim \rho_d\ ,\label{eq:M4}
\end{equation}
where $t^{'}$ and $b^{'}$ indicate the fourth generation quarks, 
and all coefficients are omitted.
The fourth generation quark mixings with the other three generations 
can also be estimated as
\begin{eqnarray}
&&|V_{ub^{'}}| \simeq |V_{t^{'}d}| \simeq |\eta_d\xi_d\rho_d|
  = \lambda^3|\rho_d| \ ,\nonumber\\
&&|V_{cb^{'}}| \simeq |V_{t^{'}s}| \simeq |\xi_d\rho_d| 
  = \lambda^2 |\rho_d|\ ,\label{eq:4mix}\\
&&|V_{tb^{'}}| \simeq |V_{t^{'}b}| \simeq |\rho_d|\ ,\nonumber
\end{eqnarray}
where we have used $\eta_d = \lambda$ and $\xi_d=\lambda^2$.
As one can see, the fourth generation quark mixing angles are directly 
related to the mass ration $m_b/m_{b^{'}}$ via $\rho_d$.
We note that the contribution of $\rho_u$ to the mixing angles 
is negligibly small compared with that of $\rho_d$.
To illustrate our implications, we substitute the lower bound of $m_{b^{'}}$,
i.e., $m_{b^{'}}>199\ \gev$ \cite{PDG}, and the central value 
$m_b=4.2\ \gev$ \cite{PDG}, into Eq. (\ref{eq:M4}).
Then we get $\rho_d < \lambda^{2}$ and
\begin{eqnarray}
&&|V_{ub^{'}}| \simeq |V_{t^{'}d}| < \lambda^5 \ ,\\
&&|V_{cb^{'}}| \simeq |V_{t^{'}s}| < \lambda^4\ ,\\
&&|V_{tb^{'}}| \simeq |V_{t^{'}b}| < \lambda^2\ .
\end{eqnarray}
It is interesting to see that $|V_{tb^{'}}|$ and $|V_{t^{'}b}|$ can be large, 
which could be measurable at the upcoming 
experiments like the LHC.

Finally, we briefly comment about CP violation.
Although, in general, the discussion of CP violation in the fourth generation 
model is very complicated, in the chiral limit of $m_{u,d,s,c}=0$, CP violation 
is described by only one simple quantity \cite{CP4}
\begin{eqnarray}
J_4={\rm Im}[V_{tb} V_{t^{'}b}^{*} V_{t^{'}b^{'}} V_{tb^{'}}^{*}]\ .
\end{eqnarray}
As discussed in Ref. \cite{soni}, the quantity may provide us useful information
about the size of CP violation.
By introducing the same phase factors discussed in Sec. III, we 
find that $J_4 \simeq 10^{-7}$.
Unfortunately, the result is much smaller than the SM Jarlskog 
invariant parameter $J_{SM}\simeq 10^{-5}$.
Hence, the fourth generation model, in our scheme, cannot be the main source of 
the CP violating phenomena, such as the baryon asymmetry of 
the Universe \cite{SM4BAU}.

\section{Summary}
We have studied the quark masses and mixing angles with the hierarchical 
FL symmetry.
We have shown that the symmetry can explain the hierarchies in the
quark masses and mixing angles at the same time.
As a result, the masses and mixing angles are closely related 
with each other in our model.
To generate light quark masses and CP violation, we have introduced 
phase factors into the mass matrices, 
and then demonstrated that our model can reproduce all experimental values.
We have also extended our scheme to the fourth generation quark model and
found the relation 
$|V_{tb^{'}}| \simeq |V_{t^{'}b}| \simeq m_b/m_{b^{'}}< \lambda^2$ 
for $m_b=4.2\ \gev$ and $m_{b^{'}}>199\ \gev$.
We have speculated that $|V_{tb^{'}}|$ and $|V_{t^{'}b}|$ could be 
measurable at the upcoming experiments like the LHC in our extended fourth 
generation quark model.
\\ 

\noindent {\bf Acknowledgments}

This work is supported in part by the National Science Council of ROC under 
Grants No. NSC-95-2112-M-007-059-MY3 and No. NSC-98-2112-M-007-008-MY3
and by the Boost Program of NTHU.

\end{document}